\documentclass[pdflatex,sn-mathphys-num]{sn-jnl}

\usepackage[compact]{titlesec}
\titlespacing{\section}{0pt}{2ex}{1ex}

\usepackage{amsmath}  
\usepackage{amsfonts} 
\usepackage[utf8]{inputenc}
\usepackage{breqn}
\usepackage{amssymb}

\DeclareMathOperator\erf{erf}

\begin{document}
	
	\title[Article Title]{A differential derivation of the Obara-Saika relation for Gaussian electron repulsion integrals}

    \author*[1,2]{\fnm{Charles C.} \sur{Forgy}}\email{ccforgy@umd.edu}

    \author[2]{\fnm{David A.} \sur{Mazziotti}}\email{damazz@uchicago.edu}

    \affil*[1]{\orgdiv{Applied Research Laboratory for Intelligence and Security}, \orgname{University of Maryland}, \orgaddress{\street{7005 52nd Ave.}, \city{College Park}, \postcode{20742}, \state{Maryland}, \country{USA}}}

	\affil[2]{\orgdiv{Department of Chemistry and The James Franck Institute}, \orgname{The University of Chicago}, \orgaddress{\street{929 E 57th St.}, \city{Chicago}, \postcode{60637}, \state{Illinois}, \country{USA}}}

	\date{\today}
	
	\abstract {The Obara-Saika (OS) method is one of the most widely used techniques in quantum chemistry for evaluating electron repulsion integrals (ERIs) via a set of recurrence relations that build higher angular momentum integrals from lower-order ones. The original derivation by Obara and Saika~\cite{Obara_1986} proceeded by directly relating integrals of differing angular momentum. In this work, we present a compact novel derivation of the OS vertical recurrence relation based solely on differential relations between Gaussian basis functions, expanding on a method suggested in earlier work. By explicitly deriving the required derivative expressions we identify all non-zero primitive terms contributing to the full ERI to develop a hierarchical formulation of the OS recursion relations. This approach has pedagogical value as a rigorous and self-contained derivation. Additionally, the resulting organization exposes independent primitive derivative quantities and may be useful for code generation and parallel implementations on modern GPU architectures.}


\maketitle
\section{Introduction}
\label{sec:intro}

Quantum chemistry methods such as ground and excited state energy estimation and molecular geometry optimization involve complex two-center four-orbital integrals that require substantial computing power to solve. Consequently, multiple methods have been proposed to simplify their evaluation by reusing as many intermediate results as possible~\cite{Obara_1986, McMurchie_1978, Boys_1960, Samu_2018}. In this work we will focus on the Obara-Saika~\cite{Obara_1986} (OS) recurrence relation used in multiple quantum chemistry packages~\cite{Lehtola_2022, Asadchev_2023, Asadchev_2023b, Asadchev_2024,  Kim_2024}. Obara and Saika originally derived their recurrence relations by relating integrals of lower angular momentum orbitals to integrals of higher angular momentum. This was primarily done with a focus on lowering the floating point operation (FLOP) count, which was the main computational bottleneck on the computers available at the time.

There has been increasing recognition that, for modern computational science, the largest bottlenecks are usually memory bandwidth and input/output (I/O) operations, not FLOPs. This has led to attempts to reformulate ERI solutions with a focus on parallelizability with the goal of leveraging the massive parallel-computation abilities of modern graphics processing units (GPUs)~\cite{Pritchard_2016,Tornai_2019, Gordon_2020, Asadchev_2023}. For example, the GPU4PySCF package~\cite{Li_2025} is designed to accelerate PySCF~\cite{Sun_2018} by calculating ERIs on GPUs using Rys quadrature~\cite{Dupuis_1976}. Similarly, Palethorpe and Barca~\cite{Palethorpe_2025} and Fujii et al.~\cite{Fujii_2025} have designed efficient implementations of ERIs on GPU architectures using the McMurchie-Davidson recurrence relation~\cite{McMurchie_1978}.

In this paper we expand on previous work~\cite{Ahlrichs_2006, Limpanuparb_2012} on a purely differential derivation of the OS recurrence relations~\footnote{Endnote (10) of Head-Gordon and Pople~\cite{Head-Gordon_1988} alludes to but does not explicitly derive the above method for the OS-VRR.} designed around differentiation with respect to nuclear coordinates. We build on earlier work by explicitly deriving the derivative relations from the basic form of Gaussian orbitals and the Boys function with an  explicit Hermite-to-Cartesian formulation, allowing us to identify all non-zero primitive terms used to calculate the full ERI. We also explore the scaling of a purely derivative based method using the General Leibniz rule and Fa\`{a} di Bruno's formula. Beyond the primary pedagogic value of a rigorous and straightforward derivation of the OS recurrence relations, this formulation also allows focus on the primitive terms that can be calculated independently in a parallel implementation. Our formulation is a particularly strong conceptual match to geometry optimization, which involves direct differentiation of ERIs with respect to nuclear coordinates. While the differential method presented will increase the FLOP count of the OS method to some extent, the savings from improved parallelizability has the potential to outweigh this cost by a substantial margin.

The remainder of this paper is arranged as follows. Sec.~\ref{sec:prev} describes electron-repulsion integrals and Gaussian orbitals and then discusses the OS method at a high level. Sec.~\ref{sec:curr_meth} composes the bulk of this paper and covers the new proposed method. We introduce the form of the differential operator in Sec.~\ref{sec:diff_op}, followed by the terms stemming from the differential operator in Sec.~\ref{sec:diff_def}. Sec.~\ref{sec:scaling} explores the scaling of the differential method and motivates the use of recurrence relations by demonstrating how pure differentiation becomes computationally unfavorable. We present an explicit differential derivation of the OS method in Sec.~\ref{sec:vert_rec_rel}, followed by concluding observations in Sec.~\ref{sec:conclusions}.

\section{Previous Methods}
\label{sec:prev}

\subsection{Electron Repulsion Integrals}
\label{sec:gauss_eris}

In the context of electronic structure calculations and gradient-based geometry optimization~\cite{Pulay_1969, Gerratt_1968, Park_2020, Park_2025, Hennefarth_2025}, integrals take the form of

\begin{equation}
	\label{eq:coulomb_int_def}
	J_{mn} = \iint \phi_m^{*}(\mathbf{r})\phi_m(\mathbf{r}) \frac{1}{|\mathbf{r} - \mathbf{r\prime}|} \phi_n^{*}(\mathbf{r\prime})  \phi_n (\mathbf{r\prime})d\textbf{r}\,d\mathbf{r\prime},
\end{equation}

and

\begin{equation}
	\label{eq:exchange_int_def}
	K_{mn} = \iint \phi_m^{*}(\mathbf{r}) \phi_n(\mathbf{r}) \frac{1}{|\mathbf{r} - \mathbf{r\prime}|} \phi_n^{*}(\mathbf{r\prime}) \phi_m(\mathbf{r\prime}) d\textbf{r}\,d\mathbf{r\prime},
\end{equation}
where $\phi_i(\mathbf{r})$ denotes orbital $i$ with coordinate vector $\mathbf{r}$. 

$J_{mn}$ is referred to as the \textit{Coulomb integral} and represents the classical electrostatic repulsion between electrons in orbitals $m$ and $n$. $K_{mn}$ is referred to as the \textit{exchange integral} and represents a purely quantum mechanical effect arising from imposing antisymmetry on the overall system~\cite{Szabo_1989, Levine_2008}. In quantum chemistry simulations, the orbitals $\phi_i$ are always purely real, and consequently the complex conjugate notation is generally omitted. Eqs.~\ref{eq:coulomb_int_def} and \ref{eq:exchange_int_def} may be combined into a single notation known as an \textit{Electron Repulsion Integral} (ERI) by labeling all four orbitals distinctly

\begin{equation}
	\label{eq:ERI_def}
	\iint \phi_m(\mathbf{r}) \phi_n(\mathbf{r}) \frac{1}{|\mathbf{r} - \mathbf{r\prime}|}\phi_k(\mathbf{r\prime}) \phi_{\ell}(\mathbf{r\prime}) d\textbf{r}\,d\mathbf{r\prime}
\end{equation}

where we have dropped the complex conjugate notation.

\subsection{Gaussian Orbitals}
\label{sec:gauss_types}

The overwhelming majority of quantum chemistry simulations use Gaussian orbitals for the basis functions $\phi_i$ in Eqs.~\ref{eq:coulomb_int_def} and \ref{eq:exchange_int_def}. There are three major types of Gaussian basis functions in common use in electronic structure calculations--either as basis functions in their own right or intermediates in calculations.

(a) Cartesian Gaussians (CG) are defined as 
\begin{equation}
\label{eq:cart_def} 
\phi_{C}(\mathbf{r}, \alpha, \mathbf{A},  n_{A_{x}}, n_{A_{y}}, n_{A_{z}}) = N  (x-A_{x})^{ n_{A_{x}}}(y-A_{y})^{ n_{A_{y}}}(z-A_{z})^{ n_{A_{z}}}e^{-\alpha(\mathbf{r}-\mathbf{A})^{2}}
\end{equation}
for normalization constant $N,$ exponential coefficient $\alpha$, and nuclear coordinate vector $\mathbf{A}$, and where $(\mathbf{r}-\mathbf{A})^{2} = (x-A_{x})^{2}+(y-A_{y})^{2}+(z-A_{z})^{2}$.
CGs are defined in terms of their total angular momentum number $n=n_{A_{x}}+ n_{A_{y}}+n_{A_{z}}$.  Thus the $n=0$ CG is the $s$-type orbital, the three $n=1$ CGs are $p$-type orbitals, and so on.

CGs obey the identities (dropping the normalization constant for the moment)
\begin{equation}
\begin{split}
\label{eq:cart_ident}
&\dfrac{\partial}{\partial A_{x}} \phi_{C}(\mathbf{r}, \alpha, \mathbf{A},  n_{A_{x}}, n_{A_{y}}, n_{A_{z}})
= 2 \alpha \phi_{C}(\mathbf{r}, \alpha, \mathbf{A},  n_{A_{x}}+1, n_{A_{y}}, n_{A_{z}})\\
&- n_{A_{x}} \phi_{C}(\mathbf{r}, \alpha, \mathbf{A},  n_{A_{x}}-1, n_{A_{y}}, n_{A_{z}}),
\end{split}
\end{equation}
and
\begin{equation}
\label{eq:cart_d_alpha}
\begin{split}
&(-1)^{n}\dfrac{\partial^{n}}{\partial \alpha^{n}} \phi_{C}(\mathbf{r}, \alpha, \mathbf{A},  n_{A_{x}}, n_{A_{y}}, n_{A_{z}}) \\
&=  \phi_{C}(\mathbf{r}, \alpha, \mathbf{A},  n_{A_{x}}+2n, n_{A_{y}}, n_{A_{z}}) 
+ \phi_{C}(\mathbf{r}, \alpha, \mathbf{A},  n_{A_{x}}, n_{A_{y}}+2n, n_{A_{z}})\\
&+ \phi_{C}(\mathbf{r}, \alpha, \mathbf{A},  n_{A_{x}}, n_{A_{y}}, n_{A_{z}}+2n).
\end{split}
\end{equation}
For $n=1$ we can also consider the relation in Eq.~\ref{eq:cart_d_alpha} as
\begin{equation}
\label{eq:cart_s_alpha}
-\dfrac{\partial}{\partial \alpha} \phi_{C}(\mathbf{r}, \alpha, \mathbf{A}, 0,0,0) =|\mathbf{r}-\mathbf{A}|^{2} \phi_{C}(\mathbf{r}, \alpha, \mathbf{A}, 0,0,0),
\end{equation}
which among other uses provides a simple method of creating angular momentum $s$-orbitals.

(b) Hermite Gaussians (HG) are defined through the differential relation 
\begin{equation}
\label{eq:her_def} 
\phi_{H}(\mathbf{r}, \alpha, \mathbf{A},  n_{A_{x}}, n_{A_{y}}, n_{A_{z}}) = N \frac{1}{(2\alpha)^{n}}  \dfrac{\partial^{n_{A_{x}}}}{\partial A_{x}^{n_{A_{x}}}}  \dfrac{\partial^{n_{A_{y}}}}{\partial A_{y}^{n_{A_{y}}}}  
\dfrac{\partial^{n_{A_{z}}}}{\partial A_{z}^{n_{A_{z}}}}  e^{-\alpha(\mathbf{r}-\mathbf{A})^{2}}.
\end{equation}
(Generally speaking, the $\tfrac{1}{(2\alpha)^{n}}$ term in Eq.~\ref{eq:her_def} is included in the normalization constant.  We write it explicitly as we shall make use of it in our derivations.) HGs obey the trivial identity
\begin{equation}
\label{eq:herm_ident}
\dfrac{\partial}{\partial A_{x}} \phi_{H}(\mathbf{r}, \alpha, \mathbf{A},  n_{A_{x}}, n_{A_{y}}, n_{A_{z}}) = 2 \alpha \, \phi_{H}(\mathbf{r}, \alpha, \mathbf{A},  n_{A_{x}}+1, n_{A_{y}}, n_{A_{z}}).
\end{equation}
Like with CGs, HGs are grouped into $s$-type, $p$-type, etc., based on total angular momentum.

(c) Solid Harmonic Gaussians (SHG) are defined as 
\begin{equation}
\label{eq:spher_def} 
\phi_{S}(\mathbf{r}, \alpha, \mathbf{A},  l, m) = N  S_{l,m}e^{-\alpha(\mathbf{r}-\mathbf{A})^{2}}
\end{equation}
where $S_{l,m}$ is a real solid harmonic defined from the real part of the spherical harmonic solutions to Laplace's Equation
\begin{equation}
\label{eq:laplace_def} 
\nabla^{2} \phi = 0. 
\end{equation}
Although SHGs (and the extremely closely related spherical harmonic Gaussians) are often used as basis functions, we shall only derive integrals in terms of CGs and HGs.  When SHGs are used as a basis set, it will be necessary to use transformations such as those given in Schlegel and Frisch~\cite{Schlegel_1995} or Reine et al.~\cite{Reine_2007} to obtain the final results.  Note that Eq.~\ref{eq:cart_s_alpha} provides a simple method of deriving the $r^{2n}$ terms used in SHGs.

Finally, we note that the $s$- and $p$-type ($n=0,1$ or $l=0,1$) Gaussians are identical for CGs, HGs, and SHGs.

Let the $(ss|ss)$ Electron Repulsion Integral (ERI) be represented as~\cite{Gill_1994, Helgaker_2000}
\begin{equation} 
\label{eq:s-ERI}
\begin{split} 
&(\phi(\mathbf{r}, \alpha, \mathbf{A},0,0,0) \phi(\mathbf{r}, \beta, \mathbf{B},0,0,0) | \phi(\mathbf{r\prime}, \gamma, \mathbf{C},0,0,0) \phi(\mathbf{r\prime}, \delta, \mathbf{D},0,0,0))\\
&=N \iint e^{-\alpha(\mathbf{r}-\mathbf{A})^{2}} e^{-\beta(\mathbf{r}-\mathbf{B})^{2}}\frac{1}{|\mathbf{r}-\mathbf{r\prime}|} e^{-\gamma(\mathbf{r\prime}-\mathbf{C})^{2}} e^{-\delta(\mathbf{r\prime}-\mathbf{D})^{2}} 
 d\mathbf{r}\,d\mathbf{r\prime},
\end{split}
\end{equation}
where $N$ is a normalization constant which will be defined later in the paper in Eq.~\ref{eq:Norm_def}.

To simplify the notation, we may also represent Eq.~\ref{eq:s-ERI} simply as
\begin{equation}
\label{eq:s-ERI_simp}
( \mathbf{A} \mathbf{B}| \mathbf{C} \mathbf{D}).
\end{equation}
Where it is necessary to specify whether an ERI is in a HG or CG basis, we shall use the notation $( \mathbf{A} \mathbf{B}| \mathbf{C} \mathbf{D})^{H}$ or $( \mathbf{A} \mathbf{B}| \mathbf{C} \mathbf{D})^{C}$, respectively.

\subsection{Obara-Saika Method}
\label{sec:OS}

In Obara and Saika's original 1986 paper~\cite{Obara_1986}, an eight term vertical recurrence relation (VRR) was presented for CGs.  Letting $a=n_{A_{i}}$ etc., the Obara-Saika VRR (OS-VRR) may be represented as
\begin{equation}
\label{eq:OS-VRR}
\begin{split}
&  (\mathbf{A}_{i,a} \mathbf{B}_{i,b} | \mathbf{C}_{i,c} \mathbf{D}_{i,d+1})^{m}= 
 (C_{i}-D_{i}) \frac{\gamma}{\gamma + \delta} (\mathbf{A}_{i,a} \mathbf{B}_{i,b}| \mathbf{C}_{i,c} \mathbf{D}_{i,d})^{m}\\ 
&+\frac{\alpha + \beta}{\alpha + \beta + \gamma + \delta} \left( \frac{\alpha A_{i}+\beta B_{i}}{\alpha+\beta}-\frac{\gamma C_{i}+\delta D_{i}}{\gamma+\delta}\right)(\mathbf{A}_{i,a} \mathbf{B}_{i,b}| \mathbf{C}_{i,c} \mathbf{D}_{i,d})^{m+1}\\
& +\frac{d}{2(\gamma + \delta)} \bigg( (\mathbf{A}_{i,a} \mathbf{B}_{i,b}| \mathbf{C}_{i,c} \mathbf{D}_{i,d-1})^{m} - \frac{\alpha+\beta}{\alpha+\beta+\gamma+\delta}(\mathbf{A}_{i,a} \mathbf{B}_{i,b}| \mathbf{C}_{i,c} \mathbf{D}_{i,d-1})^{m+1}\bigg)\\
& +\frac{c}{2(\gamma + \delta)} \bigg( (\mathbf{A}_{i,a} \mathbf{B}_{i,b}| \mathbf{C}_{i,c-1} \mathbf{D}_{i,d})^{m}- \frac{\alpha+\beta}{\alpha+\beta+\gamma+\delta}(\mathbf{A}_{i,a} \mathbf{B}_{i,b}| \mathbf{C}_{i,c-1} \mathbf{D}_{i,d})^{m+1}\bigg)\\
& + \frac{a}{2(\alpha + \beta + \gamma + \delta)}(\mathbf{A}_{i,a-1} \mathbf{B}_{i,b}| \mathbf{C}_{i,c} \mathbf{D}_{i,d})^{m+1}\\
&+ \frac{b}{2(\alpha + \beta + \gamma + \delta)}(\mathbf{A}_{i,a} \mathbf{B}_{i,b-1}| \mathbf{C}_{i,c} \mathbf{D}_{i,d})^{m+1},
\end{split}
\end{equation}
for Cartesian coordinate $i\in\{x,y,z\}$ where the $j$ and $k$ angular momentum values remain constant through all terms.  Obara and Saika's original derivation is not presented here, 
as a novel derivation will be presented in Sec.~\ref{sec:vert_rec_rel}.  We will  elaborate on the interpretation of Eq.~\ref{eq:OS-VRR} there.

\section{Current Method}
\label{sec:curr_meth}

\subsection{Differential Operators}
\label{sec:diff_op}

Our method is predicated upon representing Gaussian ERIs as two functions, $K(u)$ and $F_{n}(T)$, and their derivatives. $K(u)$ is the exponential function, i.e. $e^{u}$, and $F_{n}(T)$ is the $n$-th order Boys function defined below. The exponential term $K(u)$ and the zeroth order Boys function $F_0(T)$ are obtained by integrating Eq.~\ref{eq:s-ERI}; note that this single integral is the only one we will need, all further manipulations will be performed via differentiation and recursion.

The zeroth order Boys function is defined as~\footnote{The normalization constant of the Boys function is folded into the overall normalization constant given in Eq.~\ref{eq:Norm_def}}
\begin{equation}
\label{eq:Boys0_def}
F_{0}(T) =  \frac{\erf \sqrt{T}}{\sqrt{T}},
\end{equation}

while higher order Boys functions can be obtained via differentiation

\begin{equation}
\label{eq:Boysn_def}
F_{n}(T) =(-1)^{n} \dfrac{d^{n}}{d T^{n}} F_{0}(T).
\end{equation}

Alternatively, Boys functions may be related to one another via upward or downward recursion relations~\cite{Helgaker_2000}. Use of recursion allows calculation of a single Boys function $F_n(T)$ for a given $T$ for an entire batch of ERIs. For the purpose of demonstrating the derivation we will assume use of Eq.~\ref{eq:Boysn_def}, though we stress that in practice this method is disfavored due to numerical instability.

The arguments of the functions are as follows:
\begin{equation}
\label{eq:u_def}
u =-\frac{\alpha \beta}{\alpha+\beta}(\mathbf{A}-\mathbf{B})^{2}-\frac{\gamma \delta}{\gamma+\delta}(\mathbf{C}-\mathbf{D})^{2},
\end{equation}

and 

\begin{equation}
\label{eq:T_def}
T =\frac{(\alpha+\beta)(\gamma+\delta)}{\alpha+\beta+\gamma+\delta}
\sum_{i=x,y,z} \left( \frac{\alpha A_{i}+\beta B_{i}}{\alpha+\beta}-\frac{\gamma C_{i}+\delta D_{i}}{\gamma+\delta}\right)^{2}.
\end{equation}

Finally, the normalization constant for the target (Cartesian) integral can be expressed as:

\begin{multline}
\label{eq:Norm_def}
N(\alpha, \beta, \gamma, \delta, n_{A_{x}}, n_{A_{y}},\dots, n_{D_{z}}) = \\
8 \frac{ \alpha^{\frac{1}{2}(n_{A_{x}}+n_{A_{y}}+n_{A_{z}}+ \frac{3}{2})} \beta^{\frac{1}{2}(n_{B_{x}}+n_{B_{y}}+n_{B_{z}}+ \frac{3}{2})}  \gamma^{\frac{1}{2}(n_{C_{x}}+n_{C_{y}}+n_{C_{z}}+ \frac{3}{2})}  \delta^{\frac{1}{2}(n_{D_{x}}+n_{D_{y}}+n_{D_{z}}+ \frac{3}{2})}}{(\alpha+\beta)(\gamma+\delta)\sqrt{\alpha+\beta+\gamma+\delta}}\\
\times \sum_{i=n_{A_{x}}, n_{A_{y}},\dots, n_{D_{z}}} \frac{2^{i}}{\sqrt{(2i-1)!!}}.
\end{multline}

Where $n_{A_{x}}$ is the angular momentum in the $x$ coordinate of the orbital on center $\mathbf{A}$, and the other terms are interpreted analogously. 

Using this notation, the integrated form of Eq.~\ref{eq:s-ERI} is expressed as  
\begin{equation}
\label{eq:s-orbital-new}
N(\alpha, \beta, \gamma, \delta, 0, 0, \dots, 0) K(u) F_{0}(T).
\end{equation}

Eqs.~\ref{eq:cart_ident} and \ref{eq:herm_ident} allow us to relate higher order Gaussians to the derivatives of lower order ones with respect to the nuclear coordinates.  Note that Eq.~\ref{eq:s-ERI} is integrated with respect to $\mathbf{r}$ and $\mathbf{r\prime}$.  Thus differentiation of Eq.~\ref{eq:s-ERI} with respect to the nuclear coordinates may be applied to the integrated form in Eq.~\ref{eq:s-orbital-new}.  This allows for a conceptually simple method of relating solutions to ERIs of Gaussians possessing higher angular momentum to Eq.~\ref{eq:s-orbital-new}, as suggested by Boys in his original paper~\cite{Boys_1950} on Gaussian basis sets.

To explore this method of solving ERIs, we begin by defining a transformation operator to operate on Eq.~\ref{eq:s-orbital-new}, $\hat{C}(n, \alpha, A_{i})$. We will not explicitly reference the normalization constants in order to simplify the notation, though they are assumed to be present implicitly. The operator will represent the ERI's member Cartesian Gaussian of order $n$ as a linear combination of Hermite Gaussian functions up to order $n$

\begin{equation}
\label{eq:c-operator}
 \hat{C}(n, \alpha, A_{i})= 
    \begin{cases}
     \dfrac{n!}{2^{n}} \sum\limits_{k=0}^{\frac{n}{2}} \dfrac{1}{(\frac{n}{2}-k)!\,(2k)!} \,\dfrac{1}{\alpha^{\frac{n}{2}+k}}\dfrac{\partial^{2k}}{\partial A_{i}^{2k}}, & \text{$n$ even} \\
      \dfrac{n!}{2^{n}} \sum\limits_{k=0}^{\frac{n-1}{2}} \dfrac{1}{(\frac{n}{2}-k-\frac{1}{2})!\,(2k+1)!}\,\dfrac{1}{\alpha^{\frac{n}{2}+k+\frac{1}{2}}}\dfrac{\partial^{2k+1}}{\partial A_{i}^{2k+1}}. &  \text{$n$ odd}
    \end{cases}
\end{equation}

Eq.~\ref{eq:c-operator} can be derived for any given $n$ through iterative use of a rearranged form of Eq.~\ref{eq:cart_ident}. This rearrangement is expressed in operator notation as:
 
\begin{equation}
\label{eq:schlegel}
 \hat{C}(n+1, \alpha, A_{i})= \frac{1}{2 \alpha} \left(n\ \hat{C}(n-1, \alpha, A_{i}) +  \dfrac{\partial}{\partial A_{i}}  \hat{C}(n, \alpha, A_{i}) \right),
\end{equation}  
where $\hat{C}(-1, \alpha, A_{i}) = 0$ and $\hat{C}(0, \alpha, A_{i}) = 1$.

Finally, for any set of CGs $\{x^{2n}, y^{2n}, z^{2n}\}$ ($\{x^{2n+1}, y^{2n+1}, z^{2n+1}\}$) it is only necessary to apply Eq.~\ref{eq:c-operator} to create two of the three terms.  Eq.~\ref{eq:cart_d_alpha} and \ref{eq:cart_s_alpha} can then be used as
\begin{equation}
\label{eq:x2_y2_alpha}
 \hat{C}(2n, \alpha, A_{k}) = (-1)^{n}\dfrac{\partial^{n}}{\partial \alpha^{n}} - \hat{C}(2n, \alpha, A_{i})- \hat{C}(2n, \alpha, A_{j}).
\end{equation}
or
\begin{equation}
\label{eq:x3_y3_alpha}
\begin{split}
 \hat{C}(2n+1, \alpha, A_{k})& =  \bigg(\hat{C}(1, \alpha, A_{i})+ \hat{C}(1, \alpha, A_{j})+ \hat{C}(1, \alpha, A_{k})\bigg)(-1)^{n}\dfrac{\partial^{n}}{\partial \alpha^{n}}\\
 & - \hat{C}(2n+1, \alpha, A_{i})- \hat{C}(2n+1, \alpha, A_{j}).
 \end{split}
\end{equation}
This form can offer computational savings as  $\hat{C}(2n, \alpha, A_{k})$ ($\hat{C}(2n+1, \alpha, A_{k})$) requires the computation of a linear combination of derivatives of up to order $2n$ ($2n+1$), while the 
$\dfrac{\partial^{n}}{\partial \alpha^{n}}$ term only requires the computation of derivatives of order $n$ ($n+1$).

We note that in the context of geometry optimization via nuclear gradients, differentiation of an ERI with respect to nuclear coordinates results in the so-called Pulay force~\cite{Pulay_1969, Park_2020}. While not explicitly addressed in this paper, the relationship to Eq.~\ref{eq:c-operator} results in our method also being relevant to geometry optimization.

\subsection{Definitions of Derivatives}
\label{sec:diff_def}

We present a lemma that will be necessary later in the paper.  

\textbf{Lemma 1.} For any function $f(A_{x})$ such that $f$ is linear with respect to $A_{x}$, and any Gaussian $\phi$ where $\phi$ is a CG, HG, or SHG, the following holds:
\begin{equation}
\label{eq:f_ssss}
\begin{split}
& \dfrac{\partial^{ n_{A_{x}}}}{\partial A_{x}^{ n_{A_{x}}}} \Big(f(A_{x})  \phi(\mathbf{r}, \alpha, \mathbf{A}, 0,n_{A_{y}},n_{A_{z}} )  \Big) \\
&= n_{A_{x}}  \dfrac{\partial f(A_{x})}{\partial A_{x}} \dfrac{\partial^{n_{A_{x}}-1}}{\partial A_{x}^{n_{A_{x}}-1}}\phi(\mathbf{r}, \alpha, \mathbf{A},  0, n_{A_{y}},n_{A_{z}}) +  f(A_{x}) \dfrac{\partial^{n_{A_{x}}}}{\partial A_{x}^{n_{A_{x}}}}\phi(\mathbf{r}, \alpha, \mathbf{A},  0, n_{A_{y}},n_{A_{z}})
\end{split}
\end{equation}
\textit{Proof.} This follows directly from the Leibniz rule
\begin{equation}
\label{eq:f_leibniz}
\begin{split}
&\dfrac{\partial^{ n_{A_{x}}}}{\partial A_{x}^{ n_{A_{x}}}} \Big(f(A_{x})  \phi(\mathbf{r}, \alpha, \mathbf{A}, 0,n_{A_{y}},n_{A_{z}} )  \Big)\\
&= \sum\limits_{k=0}^{n_{A_{x}}} \binom{n_{A_{x}}}{k} \dfrac{\partial^{k} f(A_{i}) }{\partial A_{x}^{k}} \dfrac{\partial^{n_{A_{x}} - k}\phi(\mathbf{r}, \alpha, \mathbf{A},  0, n_{A_{y}},n_{A_{z}}) }{\partial A_{x}^{n_{A_{x}}-k}}\\
&= \binom{n_{A_{x}}}{n_{A_{x}}-1}  \dfrac{\partial f(A_{x})}{\partial A_{x}} \dfrac{\partial^{n_{A_{x}}-1}}{\partial A_{x}^{n_{A_{x}}-1}}\phi(\mathbf{r}, \alpha, \mathbf{A},  0, n_{A_{y}},n_{A_{z}})\\
&+  \binom{n_{A_{x}}}{n_{A_{x}}} f(A_{x}) \dfrac{\partial^{n_{A_{x}}}}{\partial A_{x}^{n_{A_{x}}}}\phi(\mathbf{r}, \alpha, \mathbf{A},  0, n_{A_{y}},n_{A_{z}})\\
&= n_{A_{x}}  \dfrac{\partial f(A_{x})}{\partial A_{x}} \dfrac{\partial^{n_{A_{x}}-1}}{\partial A_{x}^{n_{A_{x}}-1}}\phi(\mathbf{r}, \alpha, \mathbf{A},  0, n_{A_{y}},n_{A_{z}}) +  f(A_{x}) \dfrac{\partial^{n_{A_{x}}}}{\partial A_{x}^{n_{A_{x}}}}\phi(\mathbf{r}, \alpha, \mathbf{A},  0, n_{A_{y}},n_{A_{z}}),
\end{split}
\end{equation}
where the simplification in the last three lines results from the linear nature of $f(A_{x})$ causing all second and higher order derivatives to be zero.  Should $\phi$ be a HG, the identity
\begin{equation}
\label{eq:f_leibniz_her}
\begin{split}
&\dfrac{\partial^{ n_{A_{x}}}}{\partial A_{x}^{ n_{A_{x}}}} \Big(f(A_{x})  \phi_{H}(\mathbf{r}, \alpha, \mathbf{A}, 0,n_{A_{y}},n_{A_{z}} )  \Big)\\
&= \dfrac{n_{A_{x}}}{2\alpha}  \dfrac{\partial f(A_{x})}{\partial A_{x}} \phi_{H}(\mathbf{r}, \alpha, \mathbf{A},  n_{A_{x}}-1, n_{A_{y}},n_{A_{z}}) +  f(A_{x}) \phi_{H}(\mathbf{r}, \alpha, \mathbf{A},  n_{A_{x}}, n_{A_{y}},n_{A_{z}})
\end{split}
\end{equation}
follows immediately from the definition of HGs in Eq.~\ref{eq:her_def}. 

$u$ and $T$ are quadratic with respect to the coordinates, resulting in linear first derivatives, constant second derivatives, and no higher order derivatives. Hence we can represent all necessary derivatives as two first derivative vectors and two second derivative matrices.  Further, the second derivative matrices are rather sparse, as only terms of the same coordinate direction ($x$, $y$, or $z$) can be non-zero.  Briefly, the non-zero derivatives are as follows:
\begin{equation}
\label{eq:u_A}
u_{A_{i}} = -u_{B_{i}} = -2(A_{i}-B_{i}) \frac{\alpha\beta}{\alpha + \beta},
\end{equation}

\begin{equation}
\label{eq:u_C}
u_{C_{i}} = -u_{D_{i}} = -2(C_{i}-D_{i}) \frac{\gamma\delta}{\gamma + \delta},
\end{equation}

\begin{equation}
\label{eq:u_AA}
u_{A_{i}, A_{j}} = u_{B_{i}, B_{j}} = -u_{A_{i}, B_{j}} = \delta_{i,j} \frac{-2\alpha\beta}{\ \alpha \ + \ \beta},
\end{equation}
\begin{equation}
\label{eq:u_CC}
u_{C_{i}, C_{j}} = u_{D_{i}, D_{j}} = -u_{C_{i}, D_{j}} = \delta_{i,j} \frac{-2\gamma\delta}{\ \gamma \ + \ \delta},
\end{equation}
\begin{multline}
\label{eq:T_A}
\frac{1}{\alpha (\gamma + \delta)} T_{A_{i}} = \frac{1}{\beta(\gamma + \delta)} T_{B_{i}} = \frac{-1}{\gamma (\alpha + \beta)} T_{C_{i}} = \frac{-1}{\delta(\alpha + \beta)} T_{D_{i}} \\
= \frac{2}{\alpha + \beta + \gamma + \delta} \left( \frac{\alpha A_{i}+\beta B_{i}}{\alpha+\beta}-\frac{\gamma C_{i}+\delta D_{i}}{\gamma+\delta}\right),
\end{multline}
\begin{equation}
\label{eq:T_AA}
\frac{1}{\alpha^{2}} T_{A_{i}, A_{j}} = \frac{1}{\beta^{2}} T_{B_{i}, B_{j}} = \frac{1}{\alpha\beta}T_{A_{i}, B_{j}} =
 \delta_{i,j} \frac{2(\gamma+\delta)}{(\alpha+\beta)(\alpha + \beta + \gamma + \delta)},
\end{equation}
\begin{equation}
\label{eq:T_CC}
\frac{1}{\gamma^{2}} T_{C_{i}, C_{j}} = \frac{1}{\delta^{2}} T_{D_{i}, D_{j}} = \frac{1}{\gamma\delta}T_{C_{i}, D_{j}} =
 \delta_{i,j} \frac{2(\alpha+\beta)}{(\gamma+\delta)(\alpha + \beta + \gamma + \delta)},
\end{equation}
and
\begin{equation}
\label{eq:T_AC}
\frac{1}{\alpha\gamma} T_{A_{i}, C_{j}} = \frac{1}{\alpha\delta} T_{A_{i}, D_{j}} = \frac{1}{\beta\gamma}T_{B_{i}, C_{j}} = \frac{1}{\beta\delta}T_{B_{i}, D_{j}} =
 \delta_{i,j} \frac{-2}{\alpha + \beta + \gamma + \delta},
\end{equation}
where $u_{A_{i}}$ and $T_{A_{i}}$ refer to the first derivatives of $T$ and $u$ with respect to $A_{i}$, and the other terms are explained analogously.

\subsection{Structure and Scaling}
\label{sec:scaling}

Before presenting our derivation of the OS-VRR we will first motivate the computational necessity of recurrence relations in a purely differential context.  In the most na\"{\i}ve implementation of our scheme, one could follow Boys~\cite{Boys_1950} and simply apply Eq.~\ref{eq:c-operator} to Eq.~\ref{eq:s-orbital-new}, with $N$ replaced by the appropriate normalization constant.  The result will be different combinations of elements from the above mentioned vectors and matrices summed and multiplied by each-other and the Boys functions.  Using this method we have found it is possible to calculate the $L$-shell ($s-$ plus $p$-shells) at a cost of around 5,000 FLOPs. This value is somewhat higher than other methods, but is still quite reasonable for modern computer architecture.

However, beyond the $L$-shell the number of FLOPs rapidly becomes unmanageable.  To understand the reason for this, we must understand how the differential scaling works.  The scaling consists of the product of the following two rules:

(1) The General Leibniz rule: The multivariate $n$-th differential of a product of two functions has up to $2^{n}$ elements--the sum of all binomial coefficients $\sum_{k}  \binom {n} {k}$.  This is the result of application of the multivariate Leibniz (or product) rule to the product of $K(u)$ and $F_{0}(T)$, resulting in

\begin{equation}
\label{eq:multi_product}
\dfrac{\partial^{n}}{\partial A_{1} \partial A_{2} \dots \partial A_{n}} K(u) F_{0}(T) = \sum_{S} \dfrac{\partial^{|S|} K(u)}{\Pi_{i \in S} \partial A_{i}} \dfrac{\partial^{n - |S|}F_{0}(T)}{\Pi_{j \notin S} \partial A_{j}}
\end{equation}
where $S$ runs over the $2^{n}$ subsets of $\{A_{1}, A_{2}, \dots, A_{n}\}$, and $|S|$ denotes the cardinality of the set $S$.  If all $A_{i}$'s are identical, Eq.~\ref{eq:multi_product} collapses to the familiar univariate Leibniz rule, namely

\begin{equation}
\label{eq:uni_product}
\dfrac{\partial^{n}}{\partial A_{i}^{n}} K(u) F_{0}(T) = \sum\limits_{k=0}^{n} \binom{n}{k} \dfrac{\partial^{k} K(u)}{\partial A_{i}^{k}} \dfrac{\partial^{n - k}F_{0}(T)}{\partial A_{i}^{n-k}}.
\end{equation}

Most integrals in the $d$-shell and higher will have fewer than $2^{n}$ elements as not all  $A_{i}$'s will be distinct. However, as there are 12 coordinates, every shell through the $(ff|ff)$ shell contains sufficient $A_{i}$'s to contain at least one term with $n$ distinct derivatives.

(2)  Fa\`{a} di Bruno's formula:  In Eqs.~\ref{eq:multi_product} and \ref{eq:uni_product}, nothing further need be done to differentiate $K(u)$.  However, differentiating $F_{0}(T)$ requires the use of the multivariate chain rule.  This is most easily expressed using Fa\`{a} di Bruno's formula~\cite{Hardy_2006}, as follows:

\begin{equation}
\label{eq:faa_di_bruno}
\dfrac{\partial^{n}}{\partial A_{1}, \partial A_{2}, \dots, \partial A_{n} } F_{0}(T) = \sum_{S} (-1)^{|S|} F_{|S|}(T) \cdot \prod_{Q \in S_{1}, S_{2}} \dfrac{\partial^{|Q|} T}{\Pi_{j \in Q} \partial A_{j}}, \ \ \ \ n > 0
\end{equation}
where, as above, $S$ runs over all subsets of $\{A_{1}, A_{2}, \dots, A_{n}\}$.  Here, $Q\in S_{1},S_{2}$ denotes that $Q$ only consists of the $1$- and $2$-element partitions of the set $S$--the right hand side of Eq.~\ref{eq:faa_di_bruno} will be zero for any subset $S$ which cannot be further partitioned into $|S|$ 1- and/or 2-element subsets.  This is a result of the quadratic nature of $T$ in Eq.~\ref{eq:T_def}.

Fa\`{a} di Bruno's formula scales according to the Bell polynomials~\footnote{Strictly, the incomplete Bell polynomials, as all third and higher order derivatives are zero.}.  Although the Bell polynomials lack a clear asymptotic scaling, we have found that the $n$-th incomplete Bell polynomial can be very well modeled as scaling according to $0.05 e^{1.3n} \approx \mathcal{O}(e^{n})$ in the range relevant to this work.

Thus the individual integrals will scale as the product of the General Leibniz rule and Fa\`{a} di Bruno's formula, with the most difficult integrals in a given shell scaling as $2^{n} 0.05 e^{1.3n} \approx \mathcal{O}(e^{n})$, where $n$ is four times the value of the highest angular momentum orbital in the shell.

\subsection{Vertical Recurrence Relations}
\label{sec:vert_rec_rel}

Exponential scaling will result in great computational expense--if not outright intractability--for ERIs involving orbitals of higher angular momentum. Instead, let us explore methods of reusing values from lower angular momentum integrals to evaluate higher order ones.  Using the differential structure set forth in Sec.s~\ref{sec:diff_op}--\ref{sec:scaling} we may derive recurrence relations that will prove to be identical to the OS relations.  Let us examine incrementing the $(ss|ss)$ integral
by one angular momentum level.  Omitting normalization, from Eq.~\ref{eq:c-operator} we have:
\begin{equation}
\label{eq:s_2_p_1}
\begin{split}
(p_{i}s|ss) & =   \frac{1}{2\alpha}\dfrac{\partial}{\partial A_{i}} (ss|ss) \\
& =   \frac{1}{2\alpha} \dfrac{\partial}{\partial A_{i}} K(u) F_{0} (T) \\
& =\frac{1}{2\alpha} \bigg(F_{0} (T) \dfrac{\partial}{\partial A_{i}} K(u)  +   K(u) \dfrac{\partial T}{\partial A_{i}} \dfrac{d}{d T}F_{0} (T)\bigg) \\
& = \frac{1}{2\alpha}\bigg(u_{A_{i}} K(u) F_{0} (T) + T_{A_{i}} K(u)  \dfrac{d}{d T}F_{0} (T)\bigg) \\
& = \frac{1}{2\alpha} \bigg( u_{A_{i}} (ss|ss) +  T_{A_{i}} \dfrac{d}{d T} (ss|ss)\bigg).
\end{split}
\end{equation}
Let us define the operator $\hat{F}$ as 
\begin{equation}
\label{eq:f_hat_def}
\hat{F} = \frac{d}{d T}.
\end{equation}
$\hat{F}$ operates exclusively on the Boys function, and--as laid out in Eq.~\ref{eq:Boysn_def}--it functions to transform $F_{n} (T)$ into $-F_{n+1} (T)$.
The last line of Eq.~\ref{eq:s_2_p_1} can then be written as
\begin{equation}
\label{eq:s_2_p_2}
(p_{i}s|ss)  = \frac{u_{A_{i}}}{2\alpha} (ss|ss) +  \frac{T_{A_{i}}}{2\alpha} \hat{F} (ss|ss).
\end{equation}
We note that in our notation the Gaussian integral $\hat{F}^{m} (ab|cd)$ is identical to the auxiliary integral $(-1)^{m}(ab|cd)^{(m)}$ presented in the OS method. Similarly, $\hat{F}(ab|cd)$ in our notation is equivalent to $-(ab|cd)'$ in the notation of Schlegel~\cite{Schlegel_1982}.  However, the operator notation of Eq.~\ref{eq:f_hat_def} is more conducive the discussion set forth in this paper.

Continuing in this line, we may obtain expressions such as
\begin{equation}
\label{eq:s_2_pp_1}
\begin{split}
(p_{i}p_{j}|ss) &= \frac{1}{4\alpha\beta}\dfrac{\partial^{2}}{\partial A_{i} \partial B_{j}} (ss|ss)\\
&=\frac{1}{4\alpha\beta} \bigg((\delta_{i,j}u_{A_{i}, B_{j}} + u_{A_{i}} u_{B_{j}}) + (T_{A_{i}} u_{B_{j}} + u_{A_{i}} T_{B_{j}}+\delta_{i,j}T_{A_{i}, B_{j}})\hat{F}\\
&+T_{A_{i}} T_{B_{j}}\hat{F}^{2}\bigg)(ss|ss)\\
&=\frac{1}{2\beta} \bigg( u_{B_{j}} (\frac{1}{2\alpha}\big(u_{A_{i}} +T_{A_{i}})\big)(ss|ss) + \frac{\delta_{i,j}}{2\alpha}u_{A_{i}, B_{j}}(ss|ss) \\
&+ T_{B_{j}} \hat{F} (\frac{1}{2\alpha}\big(u_{A_{i}} + T_{A_{i}} \hat{F})\big)(ss|ss) +\frac{\delta_{i,j}}{2\alpha}T_{A_{i}, B_{j}}\hat{F} (ss|ss) \bigg)\\
&=\frac{1}{2\beta} \bigg( u_{B_{j}} (p_{i}s|ss) + \frac{\delta_{i,j}}{2\alpha}u_{A_{i}, B_{j}}(ss|ss) + \hat{F}[ T_{B_{j}} (p_{i}s|ss)+ \frac{\delta_{i,j}}{2\alpha}T_{A_{i}, B_{j}}(ss|ss)]\bigg)
\end{split}
\end{equation}
We may continue in this fashion to build the entire desired expression. 

Alternatively, we may invoke Eq.~\ref{eq:f_leibniz_her}. To do so, we will begin by applying the operator $\frac{1}{2\delta}\frac{\partial}{\partial D_i}$ to the unnormalized HG ERI, letting $X_{i,x}$ denote the $x$-th derivative with respect to $X_{i}$. The $j$ and $k$ values are held constant and omitted for clarity, but are manipulated in an identical manner. We will then apply Lemma 1 to systematically apply the remaining derivatives with respect to $C_i$, $B_i$, and $A_i$. This process will yield Schlegel's derivative theorem (Eq.~\ref{eq:rec_leibniz_AB}). Finally, by re-expressing the result in the CG basis using Eq.~\ref{eq:schlegel} we will arrive at the desired recurrence relation (Eq.~\ref{eq:rec_def_cart}).

\begin{equation}
\label{eq:rec_leibniz}
\begin{split}
&\dfrac{1}{2 \delta}\dfrac{\partial}{\partial D_{i}}  (\mathbf{A}_{i,a} \mathbf{B}_{i,b}| \mathbf{C}_{i,c} \mathbf{D}_{i,d})\\
&=\dfrac{1}{(2\alpha)^{a}(2\beta)^{b}(2\gamma)^{c}(2 \delta)^{d+1}}\dfrac{\partial^{a+b+c}}{\partial A_{i}^{a} \partial B_{i}^{b} \partial C_{i}^{c}}  \dfrac{\partial^{d}}{\partial D_{i}^{d}}\dfrac{\partial}{\partial D_{i}} (\mathbf{A}_{i,0} \mathbf{B}_{i,0}| \mathbf{C}_{i,0} \mathbf{D}_{i,0})\\
&=\dfrac{1}{(2\alpha)^{a}(2\beta)^{b}(2\gamma)^{c}(2 \delta)^{d+1}}\dfrac{\partial^{a+b+c}}{\partial A_{i}^{a} \partial B_{i}^{b} \partial C_{i}^{c}}\dfrac{\partial^{d}}{\partial D_{i}^{d}}  \dfrac{\partial K(u) F_{0}(T)}{\partial D_{i}} \\
&=\dfrac{1}{(2\alpha)^{a}(2\beta)^{b}(2\gamma)^{c}(2 \delta)^{d+1}}\dfrac{\partial^{a+b+c}}{\partial A_{i}^{a} \partial B_{i}^{b} \partial C_{i}^{c}}\dfrac{\partial^{d}}{\partial D_{i}^{d}}\Big(u_{D_{i}} K(u) F_{0}(T) + \hat{F}[T_{D_{i}}K(u) F_{0}(T)]\Big)\\
&=\dfrac{1}{(2\alpha)^{a}(2\beta)^{b}(2\gamma)^{c}(2 \delta)^{d+1}}\dfrac{\partial^{a+b+c}}{\partial A_{i}^{a} \partial B_{i}^{b} \partial C_{i}^{c}}\dfrac{\partial^{d}}{\partial D_{i}^{d}} \Big(u_{D_{i}} (\mathbf{A}_{i,0} \mathbf{B}_{i,0}| \mathbf{C}_{i,0} \mathbf{D}_{i,0})\\
&+ \hat{F}[T_{D_{i}} (\mathbf{A}_{i,0} \mathbf{B}_{i,0}| \mathbf{C}_{i,0} \mathbf{D}_{i,0})]\Big)\\
&=\dfrac{1}{(2\alpha)^{a}(2\beta)^{b}(2\gamma)^{c} 2 \delta}\dfrac{\partial^{a+b+c}}{\partial A_{i}^{a} \partial B_{i}^{b} \partial C_{i}^{c}} \Big(  \dfrac{d}{2 \delta} u_{D_{i}, D_{i}} (\mathbf{A}_{i,0} \mathbf{B}_{i,0}| \mathbf{C}_{i,0} \mathbf{D}_{i,d-1})\\
&+ u_{D_{i}} (\mathbf{A}_{i,0} \mathbf{B}_{i,0}| \mathbf{C}_{i,0} \mathbf{D}_{i,d})\\
&+ \hat{F}[ \dfrac{d}{2 \delta} T_{D_{i}, D_{i}} (\mathbf{A}_{i,0} \mathbf{B}_{i,0}| \mathbf{C}_{i,0} \mathbf{D}_{i,d-1})+ T_{D_{i}} (\mathbf{A}_{i,0} \mathbf{B}_{i,0}| \mathbf{C}_{i,0} \mathbf{D}_{i,d})] \Big),
\end{split}
\end{equation}
where we have made use of Lemma 1.

Applying the $\tfrac{\partial^{c}}{\partial C_{i}^{c}}$ derivative, and noting that second derivatives are not functions of the coordinates, we obtain
\begin{equation}
\label{eq:rec_leibniz_C}
\begin{split}
&\dfrac{1}{(2\alpha)^{a}(2\beta)^{b}(2\gamma)^{c} 2 \delta}\dfrac{\partial^{a+b}}{\partial A_{i}^{a} \partial B_{i}^{b}} \dfrac{\partial^{c}}{\partial C^{c}} \Big(  \dfrac{d}{2 \delta}  u_{D_{i}, D_{i}} (\mathbf{A}_{i,0} \mathbf{B}_{i,0}| \mathbf{C}_{i,0} \mathbf{D}_{i,d-1})\\
&+ u_{D_{i}} (\mathbf{A}_{i,0} \mathbf{B}_{i,0}| \mathbf{C}_{i,0} \mathbf{D}_{i,d}) + \hat{F}[\dfrac{d}{2 \delta}  T_{D_{i}, D_{i}} (\mathbf{A}_{i,0} \mathbf{B}_{i,0}| \mathbf{C}_{i,0} \mathbf{D}_{i,d-1})+ T_{D_{i}} (\mathbf{A}_{i,0} \mathbf{B}_{i,0}| \mathbf{C}_{i,0} \mathbf{D}_{i,d})] \Big)\\
&=\dfrac{1}{(2\alpha)^{a}(2\beta)^{b} 2 \delta}\dfrac{\partial^{a+b}}{\partial A_{i}^{a} \partial B_{i}^{b}} \Big(\dfrac{d}{2 \delta}  u_{D_{i}, D_{i}} (\mathbf{A}_{i,0} \mathbf{B}_{i,0}| \mathbf{C}_{i,c} \mathbf{D}_{i,d-1})\\
&+ \dfrac{c}{2 \gamma} u_{C_{i}, D_{i}} (\mathbf{A}_{i,0} \mathbf{B}_{i,0}| \mathbf{C}_{i,c-1} \mathbf{D}_{i,d}) \\
&+ u_{D_{i}} (\mathbf{A}_{i,0} \mathbf{B}_{i,0}| \mathbf{C}_{i,c} \mathbf{D}_{i,d})
+ \hat{F}[\dfrac{d}{2 \delta}  T_{D_{i}, D_{i}} (\mathbf{A}_{i,0} \mathbf{B}_{i,0}| \mathbf{C}_{i,c} \mathbf{D}_{i,d-1})\\
&+ \dfrac{c}{2 \gamma} T_{C_{i}, D_{i}} (\mathbf{A}_{i,0} \mathbf{B}_{i,0}| \mathbf{C}_{i,c-1} \mathbf{D}_{i,d}) \\
&+  T_{D_{i}} (\mathbf{A}_{i,0} \mathbf{B}_{i,0}| \mathbf{C}_{i,c} \mathbf{D}_{i,d})] \Big).
\end{split}
\end{equation}
Finally applying the derivatives $\tfrac{\partial^{b}}{\partial B_{i}^{b}}$ and $\tfrac{\partial^{a}}{\partial A_{i}^{a}}$, and noting that $u_{C_{i}}$ and $u_{D_{i}}$ are constant with respect to them, we obtain
\begin{equation}
\label{eq:rec_leibniz_AB}
\begin{split}
&\dfrac{1}{(2\alpha)^{a}(2\beta)^{b} 2 \delta}\dfrac{\partial^{a+b}}{\partial A_{i}^{a} \partial B_{i}^{b}} \Big(\dfrac{d}{2 \delta}  u_{D_{i}, D_{i}} (\mathbf{A}_{i,0} \mathbf{B}_{i,0}| \mathbf{C}_{i,c} \mathbf{D}_{i,d-1})\\
&+ \dfrac{c}{2 \gamma} u_{C_{i}, D_{i}} (\mathbf{A}_{i,0} \mathbf{B}_{i,0}| \mathbf{C}_{i,c-1} \mathbf{D}_{i,d}) + u_{D_{i}} (\mathbf{A}_{i,0} \mathbf{B}_{i,0}| \mathbf{C}_{i,c} \mathbf{D}_{i,d})\\
&+ \hat{F}[\dfrac{d}{2 \delta}  T_{D_{i}, D_{i}} (\mathbf{A}_{i,0} \mathbf{B}_{i,0}| \mathbf{C}_{i,c} \mathbf{D}_{i,d-1})+ \dfrac{c}{2 \gamma} T_{C_{i}, D_{i}} (\mathbf{A}_{i,0} \mathbf{B}_{i,0}| \mathbf{C}_{i,c-1} \mathbf{D}_{i,d})\\
&+  T_{D_{i}} (\mathbf{A}_{i,0} \mathbf{B}_{i,0}| \mathbf{C}_{i,c} \mathbf{D}_{i,d})] \Big)\\
&=\dfrac{1}{2 \delta}\Big(\dfrac{d}{2 \delta}  u_{D_{i}, D_{i}} (\mathbf{A}_{i,a} \mathbf{B}_{i,b}| \mathbf{C}_{i,c} \mathbf{D}_{i,d-1})+ \dfrac{c}{2 \gamma} u_{C_{i}, D_{i}} (\mathbf{A}_{i,a} \mathbf{B}_{i,b}| \mathbf{C}_{i,c-1} \mathbf{D}_{i,d})\\
&+ u_{D_{i}} (\mathbf{A}_{i,a} \mathbf{B}_{i,b}| \mathbf{C}_{i,c} \mathbf{D}_{i,d}) + \hat{F}[\dfrac{d}{2 \delta}   T_{D_{i}, D_{i}} (\mathbf{A}_{i,a} \mathbf{B}_{i,b}| \mathbf{C}_{i,c} \mathbf{D}_{i,d-1})\\
&+\dfrac{c}{2 \gamma} T_{C_{i}, D_{i}} (\mathbf{A}_{i,a} \mathbf{B}_{i,b}| \mathbf{C}_{i,c-1} \mathbf{D}_{i,d}) +  \dfrac{b}{2 \beta} T_{B_{i}, D_{i}} (\mathbf{A}_{i,a} \mathbf{B}_{i,b-1}| \mathbf{C}_{i,c} \mathbf{D}_{i,d})\\
&+ \dfrac{a}{2 \alpha}  T_{A_{i}, D_{i}} (\mathbf{A}_{i,a-1} \mathbf{B}_{i,b}| \mathbf{C}_{i,c} \mathbf{D}_{i,d}) +T_{D_{i}} (\mathbf{A}_{i,a} \mathbf{B}_{i,b}| \mathbf{C}_{i,c} \mathbf{D}_{i,d})]\Big).
\end{split}
\end{equation}

Eq.~\ref{eq:rec_leibniz_AB} is Schlegel's derivative theorem~\cite{Schlegel_1982}.  It is true for any HG ERI, and for CG ERIs composed solely of $s$- and $p$-type orbitals.  

To relate Eq.~\ref{eq:rec_leibniz_AB} to the the OS method,  and hence solve $d$-type and higher angular momentum CG ERIs, we re-express the CG ERI as a linear combination of HG ERIs through use of Eq.~\ref{eq:schlegel}.  Combining the formulae of Eq.~\ref{eq:schlegel} and Eq.~\ref{eq:rec_leibniz_AB}, and letting $(\mathbf{A}_{i,a} \mathbf{B}_{i,b} | \mathbf{C}_{i,c} \mathbf{D}_{i,d})^{C}$ represent the linear combination of HGs that creates the CG with angular momentum $a$ in the $i$-coordinate for center $\mathbf{A}$ etc., we obtain the OS relation
\begin{equation}
\label{eq:rec_def_cart}
\begin{split}
&  (\mathbf{A}_{i,a} \mathbf{B}_{i,b} | \mathbf{C}_{i,c} \mathbf{D}_{i,d+1})^{C}= \\
& \frac{1}{2 \delta} \Big(d\ (\frac{1}{2\delta}u_{D_{i}, D_{i}} + 1) (\mathbf{A}_{i,a} \mathbf{B}_{i,b}| \mathbf{C}_{i,c} \mathbf{D}_{i,d-1})^{C} + \frac{c}{2\gamma} u_{C_{i}, D_{i}} (\mathbf{A}_{i,a} \mathbf{B}_{i,b}| \mathbf{C}_{i,c-1} \mathbf{D}_{i,d})^{C}\\
&+   u_{D_{i}} (\mathbf{A}_{i,a} \mathbf{B}_{i,b}| \mathbf{C}_{i,c} \mathbf{D}_{i,d})^{C} + \hat{F} [ \frac{d}{2\delta} T_{D_{i}, D_{i}} (\mathbf{A}_{i,a} \mathbf{B}_{i,b}| \mathbf{C}_{i,c} \mathbf{D}_{i,d-1})^{C}\\
&+ \frac{c}{2\gamma} T_{C_{i}, D_{i}} (\mathbf{A}_{i,a} \mathbf{B}_{i,b}| \mathbf{C}_{i,c-1} \mathbf{D}_{i,d})^{C} +  \frac{b}{2\beta}T_{B_{i}, D_{i}} (\mathbf{A}_{i,a} \mathbf{B}_{i,b-1}| \mathbf{C}_{i,c} \mathbf{D}_{i,d})^{C}\\
&+ \frac{a}{2\alpha}  T_{A_{i}, D_{i}} (\mathbf{A}_{i,a-1} \mathbf{B}_{i,b}| \mathbf{C}_{i,c} \mathbf{D}_{i,d})^{C} +T_{D_{i}} (\mathbf{A}_{i,a} \mathbf{B}_{i,b}| \mathbf{C}_{i,c} \mathbf{D}_{i,d})^{C} ]\Big).
\end{split}
\end{equation}

For consistency with earlier works, we refer to the eight-term recursion relation in Eq.~\ref{eq:rec_def_cart} simply as the vertical recurrence relation (VRR). We note that Ahlrichs~\cite{Ahlrichs_2006} ultimately uses very similar differential relations, but does so without relating Hermite to Cartesian Gaussians as we do. Ahlrichs' more compact form does not allow for breaking ERIs into primitives as thoroughly as our approach.

To verify that Eq.~\ref{eq:rec_def_cart} is indeed equivalent to the OS-VRR of Eq.~\ref{eq:OS-VRR}, we substitute the explicit derivative expressions from Eqs.~\ref{eq:u_A}--\ref{eq:T_AC} and collect coefficients recalling that $\hat{F}$ in Eq.~\ref{eq:rec_def_cart} both converts from $m$ to $m+1$ in the notation of Eq.~\ref{eq:OS-VRR} and changes sign. Following the order of integrals from Eq.~\ref{eq:OS-VRR}, we have the following coefficients:

\begin{subequations}
\label{eq:equiv_proof}
\begin{align}
	\frac{u_{D_i}}{2\delta} = (C_i-D_i)\frac{\gamma}{\gamma+\delta} \qquad \text{Eq.~\ref{eq:OS-VRR} Line 1}\\
	\frac{T_{D_i}}{2\delta} = \frac{\alpha+\beta}{\alpha+\beta+\gamma+\delta}\left(\frac{\alpha A_i+\beta B_i}{\alpha+\beta} - \frac{\gamma C_i+\delta D_i}{\gamma+\delta}\right) \qquad \text{Eq.~\ref{eq:OS-VRR} Line 2}\\
	\frac{d}{2\delta}\big(\frac{1}{2\delta}u_{D_i, D_i}+1\big)= \frac{d}{2(\gamma+\delta)} \qquad	\text{Eq.~\ref{eq:OS-VRR} Line 3, Left}\\
	\frac{d}{4\delta^2}T_{D_i,D_i} = -\frac{d}{2(\gamma+\delta)}\frac{\alpha+\beta}{(\alpha+\beta+\gamma+\delta)} \qquad \text{Eq.~\ref{eq:OS-VRR} Line 3, Right}\\
	\frac{c}{4\gamma\delta}u_{C_i,D_i} = \frac{c}{2(\gamma+\delta)} \qquad	\text{Eq.~\ref{eq:OS-VRR} Line 4, Left}\\
	\frac{c}{4\gamma\delta}T_{C_i, D_i} = -\frac{c}{2(\gamma+\delta)}\frac{\alpha+\beta}{(\alpha+\beta+\gamma+\delta)}  \qquad	\text{Eq.~\ref{eq:OS-VRR} Line 4, Right}\\
	\frac{c}{4\gamma\delta}T_{A_i,D_i} = \frac{a}{2(\alpha+\beta+\gamma+\delta)} \qquad \text{Eq.~\ref{eq:OS-VRR} Line 5, Left}\\
	\frac{b}{4\beta\delta}T_{B_i,D_i} = \frac{b}{2(\alpha+\beta+\gamma+\delta)} \qquad \text{Eq.~\ref{eq:OS-VRR} Line 5, Right}
\end{align}
\end{subequations}

The left- and right-hand expressions throughout Eq.~\ref{eq:equiv_proof} are algebraically equivalent, but the left-hand forms are generally more compact, expressing each OS coefficient directly in terms of a small set of primitive derivative quantities defined in Eqs.~\ref{eq:u_A}--\ref{eq:T_AC}. Crucially, these primitives are mutually independent and need only be computed once, after which every term in the recurrence relation is obtained by multiplication and addition. This structure could be mapped onto GPU architectures, where the primitives can be evaluated concurrently across threads before being assembled into the full recurrence. While the FLOP count of this formulation is somewhat higher than the classical OS method, the ability to expose and exploit this fine-grained parallelism has the potential to more than offset this cost on modern hardware. It is therefore hoped that this differential perspective will help motivate GPU-targeted implementations of the OS-VRR in future quantum chemistry packages.

\section{Conclusions}
\label{sec:conclusions}

We have presented a novel derivation of the Obara-Saika vertical recurrence relation for Gaussian electron repulsion integrals based on a purely differential approach. Starting from the $s-$type ERI expressed in terms of the Boys function and a Gaussian exponential we recover the eight-term OS vertical recurrence relation. In the course of these derivations, we also explore the scaling of a purely differential method by applying the General Leibniz rule together with Fa\`{a} di Bruno's formula, and demonstrate how it becomes computationally unfavorable for ERIs involving higher angular momentum orbitals. While such a na\"ive implementation leads to a hierarchy that scales exponentially with the angular momentum, we demonstrate that reusing parts of the hierarchy generates a systematic set of recursion relations. These relations, while algebraically equivalent to OS, feature a different grouping of intermediates with potential advantages on modern CPU and GPU architectures

This approach offers two complementary benefits. First and most significantly, it provides a pedagogically transparent path to the OS-VRRs and--as explored in Appendix \ref{sec:hor_rec_rel}--horizontal recursion relations, grounding each term in a concrete differential operation rather than an integral identity. The arguments regarding scaling of a purely differential method provide those new to the field a clear motivation for the development of algorithms such as the OS recurrence relations. Second, the formulation makes explicit which primitive quantities—the first and second derivatives of $u$ and $T$ with respect to nuclear coordinates presented in Eqs.~\ref{eq:u_A}--\ref{eq:T_AC}—underpin the full calculation. These primitives are mutually independent and can therefore be computed in parallel in cascading series of ERI calculations, making this perspective a promising approach toward GPU-based implementations where minimizing memory latency and maximizing thread-level parallelism are the dominant concerns. While the FLOP count of this approach is somewhat higher than the classical OS method, the gains from parallelization are expected to outweigh this cost significantly for large basis sets. It is hoped that this work will inspire software developers to consider different approaches to geometry optimization and the OS relations in future quantum chemistry packages.

\bmhead{Acknowledgements}

Portions of this paper are reproduced from CF's PhD dissertation “Density Matrices: From Excitons to Molecules,” (University of Chicago, 2017). 

\section*{Declarations}

\begin{itemize}
\item Conflict of interest/Competing interests: The authors declare no conflict of interests.
\end{itemize}

\begin{appendices}

\section{Combining Integrals on a Single Center}
\label{sec:single_center}
In certain cases, we may find it more convenient to refer to all integrals of a certain angular momentum level on a given center, such as $(d_{ij}s|ss)$ or $(f_{ijk}s|ss)$.  To do so, let us first define the expanded Kronecker delta function, $\delta_{i_{1},i_{2}, \dots, i_{m}}$, to operate on the set $\{i_{1}, i_{2}, \dots, i_{m}\}$, where $m$ is an even integer:
\begin{equation}
\label{eq:del_power}
\delta_{i_{1},i_{2}, \dots, i_{m}}= \delta_{i_{1}, i_{2}}\dots\delta_{i_{m-1}, i_{m}} + \delta_{i_{1}, i_{3}}\dots\delta_{i_{m-1}, i_{m}} + \dots + \delta_{i_{1}, i_{m}}\dots\delta_{i_{2}, i_{m-1}},     \text{m even},
\end{equation}
where $\delta_{i_{1},i_{2}, \dots, i_{m}}$ contains the $\frac{1}{(m/2)!} \frac{m!}{2^{m/2}}$ 2-member partitions of $\{i_{1}, i_{s}, \dots, i_{m}\}$. For $m=0$ (the null set), Eq.~\ref{eq:del_power} is defined as $1$, while for $m=2$ Eq.~\ref{eq:del_power} collapses to the standard Kronecker delta function. Our definition of the expanded Kronecker function serves the same purpose as the multiset partitions in Hardy~\cite{Hardy_2006}. Using Eq.~\ref{eq:del_power} we may recast Eq.~\ref{eq:c-operator} as
\begin{equation}
\begin{split}
\label{eq:b-operator}
 &\hat{B}(n_{A_{x}}, n_{A_{y}}, n_{A_{z}}, \alpha, A)=\\
    &\begin{cases}
    \sum\limits_{k=0}^{\frac{n}{2}} \dfrac{1}{(2 \alpha)^{\frac{n}{2} + k}} \sum\limits_{t,u,v,w \in S}^{Q} \left( (\delta_{i_{1}=t, \dots, i_{n-2k}=u}) \dfrac{\partial^{2k}}{\prod_{v,w \neq t,u} \partial A_{v} \partial A_{w}} \right),
    & \text{$n$ even} \\
      \sum\limits_{k=0}^{\frac{n-1}{2}} \dfrac{1}{(2 \alpha)^{\frac{n}{2} + k+\frac{1}{2}}} \sum\limits_{t,u,v \in S}^{Q} \left( (\delta_{i_{1}=t, \dots, i_{n-2k-1}=u}) \dfrac{\partial^{2k+1}}{\prod_{v \neq t,u} \partial A_{v}} \right), 
      &  \text{$n$ odd}
    \end{cases}
\end{split}
\end{equation} 
where $n = n_{A_{x}} +  n_{A_{y}} + n_{A_{z}}$, $S$ is the set of $n_{A_{x}}$ $x$'s, $n_{A_{y}}$ $y$'s, and $n_{A_{z}}$ $z$'s, and $Q$ is the number of distinct partitions of $S$ into $2n-k$ ($2n-k-1$) expanded Kronecker delta functions multiplied by a derivative of order $2k$ ($2k+1$). 

\section{Horizontal Recurrence Relations}
\label{sec:hor_rec_rel}

To complete a set of ERIs, our method can be combined with existing horizontal recursion relations (HRR)~\cite{Head-Gordon_1988, Ryu_1991} using methods such as those proposed by Tornai et al.~\cite{Tornai_2019}. The HRRs may be derived from our methodology as well. Using Eq.~\ref{eq:s_2_p_2} 

\begin{equation}
\label{eq:HRR}
\begin{split}
(sp_i|ss)-(p_is|ss)  &= \frac{u_{B_{i}}}{2\beta} (ss|ss) +  \frac{T_{B_{i}}}{2\beta} \hat{F} (ss|ss)\\
&-\Big(\frac{u_{A_{i}}}{2\alpha} (ss|ss) +  \frac{T_{A_{i}}}{2\alpha}\hat{F} (ss|ss)\Big)\\
&=\Big(\frac{u_{B_{i}}}{2\beta} - \frac{u_{A_{i}}}{2\alpha} \Big)(ss|ss) + \Big(\frac{T_{B_{i}}}{2\beta}- \frac{T_{A_{i}}}{2\alpha}\Big)\hat{F} (ss|ss)\\
&=\Big(\frac{u_{B_{i}}}{2\beta} - \frac{u_{A_{i}}}{2\alpha} \Big)(ss|ss),
\end{split}
\end{equation}

where we have used the fact that by Eq.~\ref{eq:T_A} $\frac{T_{B_{i}}}{2\beta}- \frac{T_{A_{i}}}{2\alpha}=0$. Rearranging Eq.~\ref{eq:HRR} gives us the familiar HRR

\begin{equation}
    \label{eq:HRR_simp}
    (sp_i|ss) = (p_{i}s|ss)+\Big(\frac{u_{B_{i}}}{2\beta} - \frac{u_{A_{i}}}{2\alpha} \Big)(ss|ss).
\end{equation}

The term $\frac{u_{B_{i}}}{2\beta} - \frac{u_{A_{i}}}{2\alpha}$ can be simplified to $(A_i-B_i)$, though in our design this is not necessary as our algorithm will have precomputed $u_{A_i}$ and $u_{B_i}$ and need only reference their values.

$s-$ and $p-$type orbitals are identical for CGs and HGs, making Eq.~\ref{eq:HRR_simp} a conceptually appealing method of understanding HRRs for lower angular momenta integrals. For $d-$shell and higher where HGs and CGs differ, Eq.~\ref{eq:HRR_simp} could be combined with Eq.~\ref{eq:c-operator} to transform from HGs to CGs, but in practice it is mathematically far less unwieldy to derive it by subtracting the OS-VRRs 

\begin{equation}
(\mathbf{A}_{i,a+1} \mathbf{B}_{i,b} | \mathbf{C}_{i,c} \mathbf{D}_{i,d})^{C}-(\mathbf{A}_{i,a} \mathbf{B}_{i,b+1} | \mathbf{C}_{i,c} \mathbf{D}_{i,d})^{C}    
\end{equation}

as done in the original paper~\cite{Head-Gordon_1988}.

Building on sources such as Head-Gordon and Pople~\cite{Head-Gordon_1988} and Hamilton and Schaefer~\cite{Hamilton_1991}, we suggest a simple linear relation (if not a true HRR) to transform between the angular momentum from center $\mathbf{A}$ or $\mathbf{B}$ to $\mathbf{C}$ or $\mathbf{D}$, albeit at a higher cost than Eq.~\ref{eq:HRR_simp}. Using the same logic as above, we have

\begin{equation}
\label{eq:HRR_A_C}
\begin{split}
(ss|p_is)-(p_{i}s|ss)  &= \frac{u_{C_{i}}}{2\gamma} (ss|ss) +  \frac{T_{C_{i}}}{2\gamma} \hat{F} (ss|ss)\\
&-\Big(\frac{u_{A_{i}}}{2\alpha} (ss|ss) +  \frac{T_{A_{i}}}{2\alpha}\hat{F} (ss|ss)\Big)\\
&=\Big(\frac{u_{C_{i}}}{2\gamma} - \frac{u_{A_{i}}}{2\alpha} \Big)(ss|ss) + \Big(\frac{T_{C_{i}}}{2\gamma}- \frac{T_{A_{i}}}{2\alpha}\Big)\hat{F} (ss|ss).
\end{split}
\end{equation}

While perhaps not novel in principle, this form makes the relation explicit within our differential framework. In contrast with Eq.~\ref{eq:HRR}, the term $\frac{T_{C_{i}}}{2\gamma}- \frac{T_{A_{i}}}{2\alpha}$ no longer simplifies to zero except when $\alpha+\beta=\gamma+\delta$. The dependence on higher powers of $\hat{F}^n(ss|ss)$ results in Eq.~\ref{eq:HRR_A_C} being more similar to a hybrid between a VRR and an HRR rather than a true HRR. Nevertheless, given that the derivative terms from Eqs.~\ref{eq:u_A}--\ref{eq:T_AC} and all necessary $\hat{F}^n(ss|ss)$ terms have been precomputed, Eq.~\ref{eq:HRR_A_C} is still computationally inexpensive in our scheme. Eq.~\ref{eq:HRR_A_C} is more compact in its primitive form, though not directly comparable to the the HRRs proposed previously~\cite{Hamilton_1991} to move angular momentum from the first pair of coordinates to the second pair. It should be noted that the dependence on $\hat{F}(ss|ss)$ results in it not being easily amenable to application post-contraction like a true HRR. Nonetheless, Eq.~\ref{eq:HRR_A_C} does help build our method from the basis of a handful of primitives and suggest a path to a method for ERI computation that places the highest priority on parallelization rather than overall FLOPs.

\end{appendices}

\bibliography{main}

\end{document}